\documentclass[aps,prx,reprint,superscriptaddress,amsmath,amssymb]{revtex4-2}

\usepackage{graphicx}  
\usepackage{dcolumn}   
\usepackage{bm}        
\usepackage{hyperref}  
\usepackage{physics}   
\usepackage[utf8]{inputenc}
\usepackage{siunitx}   
\usepackage{float}        
\usepackage{subcaption}   
\usepackage{booktabs}

\begin{document}
	
	\title{Approaching the Quantum Limit of Optical Rotatory Dispersion:\\ From First-Principles to Single-Photon Monochromators}
	
	\author{Xianglun Ma}
	\thanks{These authors contributed equally to this work.}
	\affiliation{School of Physics, Xidian University, Xi'an, 710071, China}
	
	\author{Junyan Zhu}
	\thanks{These authors contributed equally to this work.}
	\affiliation{School of Physics, Xidian University, Xi'an, 710071, China} 
	
	\author{Fang Lu}
	\email{lufang11@163.com}
	\affiliation{School of Physics, Xidian University, Xi'an, 710071, China}
	
	\begin{abstract}
		Optical rotatory dispersion (ORD) in chiral media, classically demonstrated as the ``sweet monochromator,'' provides a robust mechanism for liquid-tunable spectral filtering. However, the ultimate physical boundaries governing its spectral bandwidth remain fundamentally unexplored beyond classical electromagnetic theory. Here, we present a comprehensive framework bridging macroscopic optical filtering with the quantum electrodynamic (QED) limits of chiral light-matter interaction. Using time-dependent density functional theory (TD-DFT) combined with Boltzmann conformational averaging, we accurately compute the ORD curves of representative saccharide systems, revealing that the theoretical minimum bandwidth is intrinsically tied to the anomalous dispersion (Cotton effect) near the molecular absorption band. While our macroscopic experiments demonstrate a classical bandwidth limit of $\sim\SI{20}{\nano\meter}$ due to heterogeneous broadening and instrumental constraints, our first-principles calculations predict achievable sub-nanometer bandwidths utilizing visible-absorbing chiral molecules. Furthermore, we extrapolate this framework to the strict quantum limit, proposing a single-photon QED architecture operating at ultra-low temperatures ($\sim\SI{10}{\milli\kelvin}$). In this regime, the spectral purity is constrained solely by the natural lifetime of the molecular excited state, governed by the Heisenberg uncertainty principle. This work establishes the ultimate theoretical criteria for chiral optical filters and pioneers the concept of the ``quantum monochromator'' for ultrasensitive chiral spectroscopy and quantum information processing.
	\end{abstract}
	
	\maketitle
	
	
	\section{Introduction}
	\label{sec:introduction}
	
	Optical rotatory dispersion (ORD) has long served as a fundamental tool for probing the electronic structure of chiral molecules. Exploiting the wavelength-dependent rotation of linearly polarized light, chiral media such as aqueous sugar solutions can function as prism-free, liquid-tunable spectral filters, classically demonstrated as the ``sweet monochromator.'' However, the performance limits of such devices have traditionally been analyzed strictly within the framework of classical electromagnetic theory. This macroscopic perspective leaves the physical origins of their ultimate spectral bandwidth limit largely unexplored at the quantum level, and a systematic theoretical framework to predict and optimize this bandwidth from molecular first principles remains absent.
	
	Fundamentally, the ultimate limit of the ORD bandwidth is inherently a quantum electrodynamics (QED) problem. As the filtering mechanism approaches the molecular absorption band edge, classical normal dispersion models fail. Dictated by the Kramers-Kronig relations, the anomalous dispersion near the absorption peak---known as the Cotton effect---indicates that the minimum achievable bandwidth is fundamentally constrained by the finite lifetime of molecular excited states and vacuum fluctuations, governed by the Heisenberg energy-time uncertainty principle. Despite this clear physical mechanism, systematic studies utilizing first-principles calculations, such as Time-Dependent Density Functional Theory (TD-DFT), to quantitatively estimate these theoretical limits and guide the design of macroscopic filter devices are exceedingly scarce.
	
	Furthermore, existing experiments predominantly operate in the macroscopic, classical regime, where the true quantum limit is masked by the statistical averaging of large photon numbers. When transmitted light is severely attenuated, the classical Malus's law description breaks down. In this single-photon limit, the coherent energy exchange between photon wave packets and chiral molecules must be described by QED models, such as the Jaynes-Cummings model. Accessing this regime requires extreme suppression of thermal decoherence and Doppler broadening, combined with high-efficiency single-photon detection---a frontier that has yet to be integrated into the study of optical rotatory filtering.
	
	In this paper, we propose a comprehensive theoretical and experimental framework bridging the classical, semiclassical, and quantum limits to determine the ultimate spectral bandwidth of chiral rotatory monochromators. We first employ TD-DFT, combined with solvation models and Boltzmann conformational averaging, to compute the ORD curves of representative systems (sucrose, fructose, and anthocyanins), predicting achievable bandwidths down to the sub-nanometer scale. We further establish a quantum-limit architecture utilizing ultra-low temperature environments ($\sim\SI{10}{\milli\kelvin}$) and superconducting nanowire single-photon detectors (SNSPDs) to access sub-natural linewidth spectra. By elucidating the dynamics of single-photon chiral coupling, this work establishes a first-principles criterion for the ultimate performance boundaries of optical rotatory filters, paving the way for the development of next-generation ``quantum monochromators.''
	
	\section{Theoretical Framework and Computational Results}
	\label{sec:theory}
	
	\section{Theoretical Framework and Computational Results}
	
	The ``sweet monochromator'' is fundamentally a macroscopic manifestation of the interaction between light and a chiral molecular medium. At the level of classical electromagnetism, this interaction is described by Maxwell's equations and the constitutive relations of the material: chiral molecules exhibit distinct refractive indices for left- and right-circularly polarized light, thereby inducing a rotation of the polarization plane of linearly polarized light. However, this classical picture fails to address a fundamental question: does a physical limit exist for the spectral bandwidth of such a filter, and what determines this limit? To answer this, one must resort to the microscopic theory of light-matter interaction. Within the framework of QED, the interaction between a molecule and the optical field is not composed of three isolated physical effects---optical rotation, dispersion, and absorption---but rather represents projections of the same physical essence onto different degrees of freedom. They are unified within the molecular complex polarizability tensor \(\widetilde{\alpha}\), whose response function can be derived from first principles via the Kubo formula of linear response theory. Briefly: absorption is determined by the imaginary part, dispersion is described by the real part, and optical rotation is a specific manifestation of dispersion on chiral structures.
	
	We divide the operational range of the system into two typical regimes. In the weak coupling regime, the incident light frequency is far from the electronic excited-state absorption peak, and the energy exchange between the molecule and the light field is negligible. This regime is dominated by normal dispersion, where the specific rotation varies smoothly with wavelength. In the strong coupling regime, as the light frequency approaches the absorption peak, resonant energy exchange occurs, leading to a drastic enhancement in absorption. The conventional dispersion description breaks down, and the specific rotation undergoes drastic changes accompanied by a sign reversal within a narrow spectral window---the Cotton effect (anomalous dispersion).
	
	Our core assertion is that an ultra-narrow spectral bandwidth cannot be attained in the smooth normal dispersion region, but must inevitably emerge in the transition zone where the Cotton effect occurs. In this regime, the derivative \(\mathrm{d}\theta/\mathrm{d}\lambda\) is maximized, allowing an infinitesimal wavelength mismatch to generate a significant difference in polarization rotation, which is subsequently effectively blocked by the analyzer.
	
	The spectral bandwidth of an optical rotatory monochromator (full width at half maximum \(\Delta\lambda\)) is primarily determined by the competition between the steepness of the ORD and the natural linewidth of the molecular excited state. The bandwidth is jointly determined by the first derivative of the specific rotation and the absorption peak bandwidth. The molecular absorption bandwidth itself is constrained by the Heisenberg energy-time uncertainty relation:
	
	\begin{equation}
		\Gamma \cdot \tau \sim \hbar,
		\label{eq:heisenberg}
	\end{equation}
	
	where \(\Gamma\) is the energy level width and \(\tau\) is the excited-state lifetime. In room-temperature solutions, collisions and thermal motion broaden the linewidth, masking this quantum limit. However, if the system is pushed to the quantum limit---an isolated chiral molecule in an ultra-low temperature environment (\(\sim\SI{10}{\milli\kelvin}\))---the absorption linewidth can reach the order of \(\SI{e-4}{\nano\meter}\) or even lower. Thus, the narrowest achievable spectral bandwidth is determined by an optimal balance between the slope of the ORD in the Cotton effect region and the natural linewidth of the excited state, corresponding to the minimum of the transmitted light intensity in experiments.
	
	To investigate this, we selected sucrose solutions and F60 high-fructose corn syrup as model systems and employed TD-DFT to calculate their ORD curves. The calculations comprehensively accounted for conformational diversity. For sucrose, we considered its dominant conformation in aqueous solution, consistent with previous studies. For fructose (fructose:glucose mass ratio 6:4), we independently calculated the contributions of four fructose tautomers (\(\alpha/\beta\)-fructopyranose, \(\alpha/\beta\)-fructofuranose) and two glucose anomers (\(\alpha/\beta\)-glucopyranose), followed by ensemble averaging based on the Boltzmann distribution at room temperature (\SI{293}{\kelvin}). The Gibbs free energy and population fraction of each conformation are detailed in Appendix~\ref{app:methods}. The calculated specific optical rotations are compared with experimental standard values in Table~\ref{tab:specific_rotation}: the error for sucrose at \SI{589}{\nano\meter} is approximately \SI{18}{\percent}, validating the reliability of our computational methodology. For the fructose/glucose mixture, the ensemble-averaged error is within \SI{15}{\percent}, which is acceptable given the complexity of the conformational space.
	
	The ORD curves of the two substances present a stark contrast. Sucrose's ORD remains positive and increases monotonically with decreasing wavelength, surging in the near-ultraviolet (UV). Conversely, fructose's ORD sign in the long-wavelength region is negative. This yields the first experimentally testable prediction: under identical conditions, the color cycle direction of the sucrose solution will be opposite to that of the fructose syrup, and the spectral bandwidth of the latter will be systematically broader.
	
	Crucially, the first derivatives of the ORD curves for both sugars, \(\mathrm{d}[\alpha]/\mathrm{d}\lambda\), decrease monotonically from the visible to the near-UV region. This implies that within the visible spectrum, the narrowest spectral bandwidth can only be realized at the boundary between visible and near-UV light. To attain a truly minimum bandwidth, the operational wavelength must be pushed into the UV region. However, the absorption peaks of common sugars are in the deep UV (\(<\SI{200}{\nano\meter}\)). A natural corollary follows: if natural chiral molecules with absorption peaks in the visible spectrum can be identified, the bandwidth can be compressed to sub-nanometer scales. Anthocyanin molecules (e.g., cyanidin-3-O-glucoside) are ideal candidates; their absorption peaks fall within the visible region, enabling the ``sweet monochromator'' to genuinely approach its quantum bandwidth limit.
	
	We conducted independent ORD calculations for each molecular component. All calculations were performed at the CAM-B3LYP/aug-cc-pVTZ level, utilizing the SMD implicit solvation model for water. The calculated specific optical rotation \([\alpha]_{\lambda}\) as a function of wavelength for each component shows that deviations for the majority of conformations are within \SI{15}{\percent}, validating the method.
	
	A prominent common feature emerges: the first derivatives for all saccharide components are substantially larger in the near-UV and decrease monotonically toward the visible spectrum. \(\beta\)-D-fructopyranose exhibits the largest first derivative value, implying its potential to achieve a narrower bandwidth.
	
	The macroscopic optical rotation is the weighted average of all conformations under thermal equilibrium. Based on the Gibbs free energies, population fractions were derived according to the Boltzmann distribution at \SI{293}{\kelvin}. Fructose exists predominantly as \(\beta\)-D-fructofuranose and \(\beta\)-D-fructopyranose (totaling \(\sim\SI{97}{\percent}\)); the \(\alpha\) anomers are negligible. Glucose exhibits a slight predominance of the \(\beta\) anomer (\(\sim\SI{53}{\percent}\)), with the \(\alpha\) form accounting for the remaining \(\sim\SI{47}{\percent}\).
	
	By linearly superimposing the single-component curves according to their Boltzmann weights, the macroscopic ORD curves for the solutions were obtained. The calculation error for sucrose (\(\sim\SI{18}{\percent}\)) is attributable to intrinsic DFT limitations, such as systematic deviations in excited-state energies and the implicit solvation model, rather than conformational sampling issues, given the dominant single-conformation character of sucrose in aqueous solution. The error for the fructose/glucose mixture is within \SI{15}{\percent}. The ORD curves reveal fundamental differences: sucrose's optical rotation remains positive and monotonically increases, while fructose undergoes a sign inversion as wavelength decreases. This provides a testable prediction: the color cycle directions of the two solutions should be opposite. This prediction is directly corroborated by experimental observations in Section~\ref{sec:experiment}.
	
	\section{Experimental Demonstration}
	\label{sec:experiment}
	
	To systematically investigate the tuning capability of the optical rotatory dispersion (ORD) effect on spectral bandwidth, we constructed a coaxial linear polarization monochromator. The optical train consists sequentially of a high-stability halogen tungsten lamp, a high-extinction polarizer, a temperature-jacketed sample cell, a rotatable analyzer equipped with a precision vernier, and a high-resolution miniature spectrometer (USB2000+UV-VIS). To decouple the instrumental response baseline, the dark background spectrum and the blank tube transmission spectrum were recorded synchronously during data acquisition. The intrinsic transmittance spectrum of the sample was then extracted via full-spectrum point-by-point normalization:
	\begin{equation}
		T_{\text{norm}}(\lambda) = \frac{I_{\text{sample}}(\lambda) - I_{\text{dark}}(\lambda)}{I_{\text{blank}}(\lambda) - I_{\text{dark}}(\lambda)}
		\label{eq:normalization}
	\end{equation}
	This normalization protocol rigorously eliminates multiplicative distortions, such as non-uniform source emission and nonlinear detector response, while simultaneously subtracting additive thermal noise. The theoretical transmittance of the system, governed by polarization interference and the Drude ORD model, is formulated as:
	\begin{equation}
		T(\lambda) = \sin^2\!\left( \theta + \frac{K C L \lambda^2}{\lambda^2 - \lambda_0^2} \right) e^{-\alpha L} (1 - d)
		\label{eq:transmittance}
	\end{equation}
	where $\theta$ is the nominal angle between the polarizer and analyzer, $C$ is the mass concentration, $L$ is the geometric path length, $K$ is the intrinsic ORD constant of the molecule, $\lambda_0$ is the near-UV resonance wavelength, $\alpha$ is the full-spectrum dissipation coefficient accounting for scattering and non-intrinsic absorption, and $d$ represents the system leakage factor induced by window birefringence and imperfect extinction ratios.
	
	To comprehensively quantify the filtering characteristics, four independent variable groups were established: (1) Angular Variation: $\theta$ was tuned in increments at fixed $C$ (g/mL) and $L$ (cm). (2) Concentration Variation: Solutions of sucrose and fructose were prepared from 0.20 to 0.60 g/mL (0.10 g/mL steps) for both 10 cm and 20 cm. (3) Path Length Variation: Cross-comparisons between 10 cm and 20 cm sample tubes were conducted, with Eq.~(\ref{eq:normalization}) recalibrated after each swap. (4) Chiral Medium Variation: Dextrorotatory sucrose, levorotatory fructose, and fructose were compared to test the structural dependence of bandwidth selection. All spectra underwent automated background subtraction and normalization via Python. The intrinsic full width at half maximum (FWHM), representing the output bandwidth, was extracted via half-maximum threshold searching of the interference envelopes.
	
\begin{figure*}[htbp]
		\centering
		\begin{subfigure}[b]{0.24\textwidth}
			\centering
			\includegraphics[width=\linewidth]{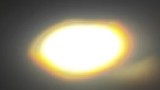}
			\caption{yellow}
		\end{subfigure}\hfill
		\begin{subfigure}[b]{0.24\textwidth}
			\centering
			\includegraphics[width=\linewidth]{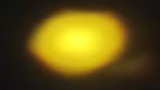}
			\caption{orange}
		\end{subfigure}\hfill
		\begin{subfigure}[b]{0.24\textwidth}
			\centering
			\includegraphics[width=\linewidth]{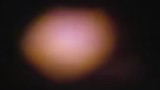}
			\caption{red}
		\end{subfigure}\hfill
		\begin{subfigure}[b]{0.24\textwidth}
			\centering
			\includegraphics[width=\linewidth]{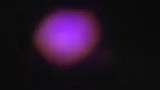}
			\caption{purple}
		\end{subfigure}
		
		\vspace{10pt} 
		
		\begin{subfigure}[b]{0.24\textwidth}
			\centering
			\includegraphics[width=\linewidth]{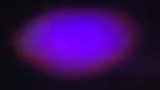}
			\caption{dark blue}
		\end{subfigure}\hfill
		\begin{subfigure}[b]{0.24\textwidth}
			\centering
			\includegraphics[width=\linewidth]{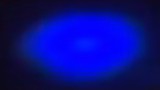}
			\caption{blue}
		\end{subfigure}\hfill
		\begin{subfigure}[b]{0.24\textwidth}
			\centering
			\includegraphics[width=\linewidth]{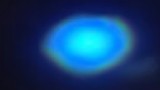}
			\caption{cyan}
		\end{subfigure}\hfill
		\begin{subfigure}[b]{0.24\textwidth}
			\centering
			\includegraphics[width=\linewidth]{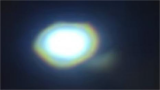}
			\caption{back to yellow}
		\end{subfigure}
		
		\caption{Visual color cycle observation of sucrose. When the polarizer and analyzer are parallel, the transmitted light appears yellowish. Rotating the analyzer clockwise, the continuous color sequence progresses through yellow, orange, red, purple, dark blue, blue, cyan, and back to yellow (a-h). This continuous sequence perfectly illustrates the wavelength-dependent polarization rotation induced by the optical rotatory dispersion (ORD) of the chiral medium. Note: For fructose (not pictured), the order of the color cycle is completely reversed, and the overall brightness is noticeably lower, consistent with the sign difference in their ORD curves derived in Section~II.}
		\label{fig:visual_color}
	\end{figure*}

	For sucrose solution, the phenomenon of transmitted light is shown in Fig.~\ref{fig:visual_color}.
	
	Several anomalous color manifestations were observed: 
	(1) Blue light appears relatively ``pure'' while violet light is almost unattainable---this is attributed to the low spectral energy of halogen lamps in the violet region, poor polarizer performance at short wavelengths, and low human eye photopic efficiency at 400 nm. 
	(2) The red region exhibits a pinkish tint due to a yellowish background from trace impurities. 
	(3) Green light is inconspicuous because the yellow background skews the output toward yellow-green. 
	(4) When only violet light is nominally received, the bandwidth paradoxically broadens because light of adjacent wavelengths arrives at the output simultaneously due to the extremely short rotatory period.
	
	\begin{figure*}[htbp]
		\centering
		\begin{subfigure}[b]{0.48\linewidth}
			\centering
			\includegraphics[width=\linewidth, height=5.5cm]{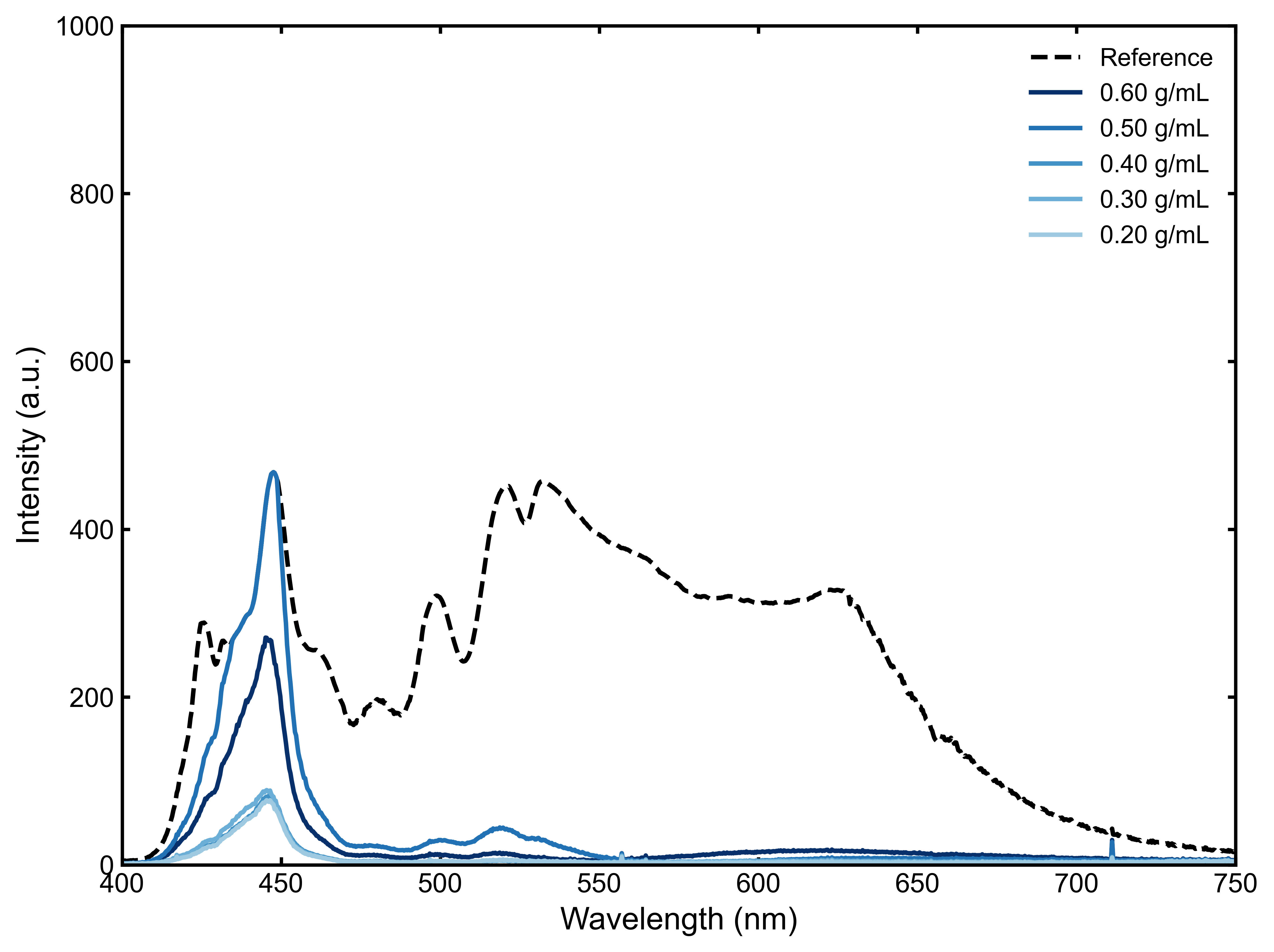}
			\caption{Raw transmitted intensity}
			\label{fig:raw}
		\end{subfigure}\hfill
		\begin{subfigure}[b]{0.48\linewidth}
			\centering
			\includegraphics[width=\linewidth, height=5.5cm]{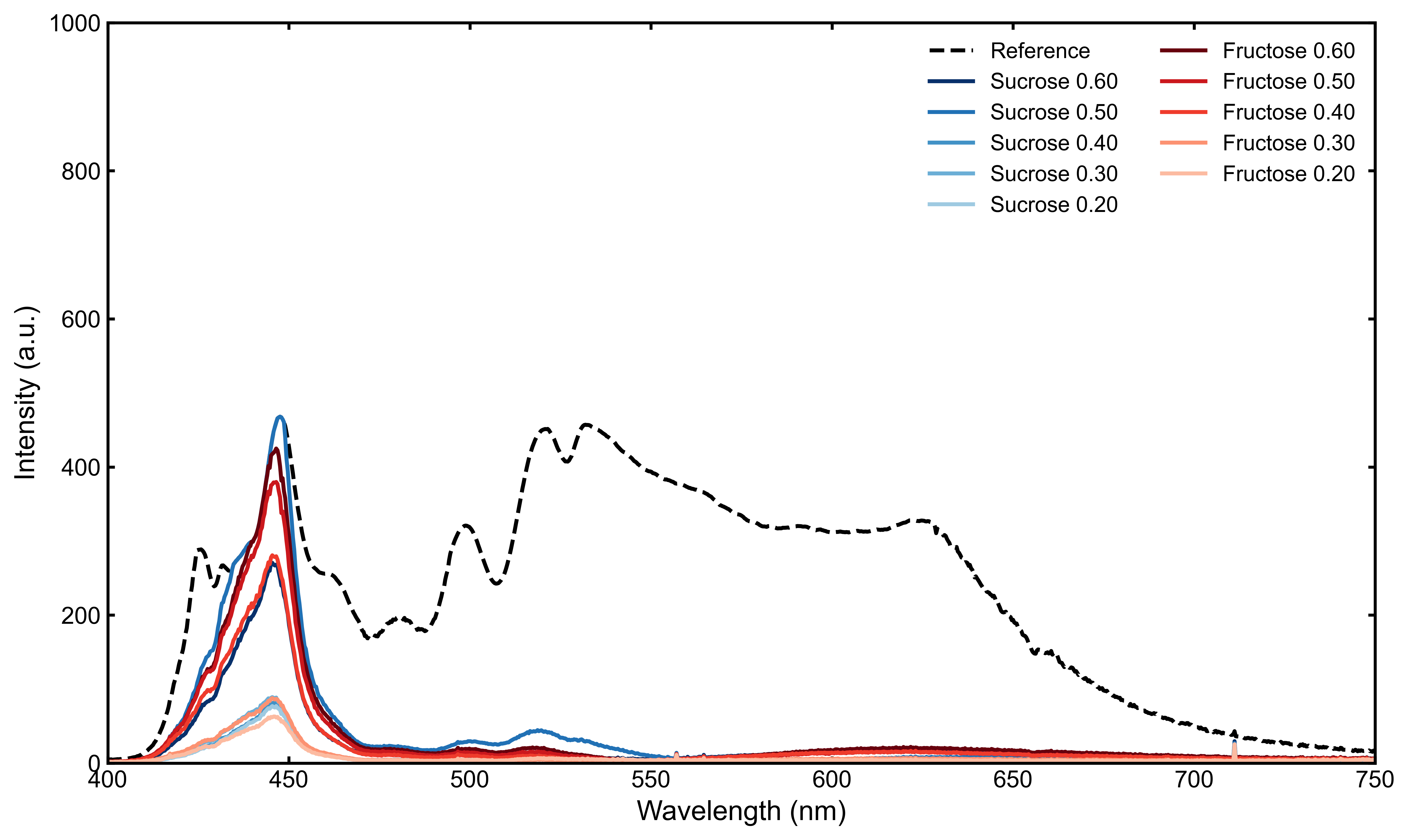}
			\caption{Normalized transmittance}
			\label{fig:norm}
		\end{subfigure}
		\caption{Figure \ref{fig:spectra_conc}(a) illustrates the raw transmitted intensity spectra under varying sucrose concentrations. While the sharp transmission peaks near 450 nm phenomenologically confirm the narrow-band filtering effect, their true geometric profiles are severely distorted by the inherent emission peaks of the halogen lamp. Figure \ref{fig:spectra_conc}(b) displays the corresponding normalized transmittance after data cleansing. The removal of baseline fluctuations reveals ideal bell-shaped interference envelopes dictated by the $\sin^2$ phase term. Notably, for $0.50$\,g/mL, a plateau reaching the physical limit is observed at the central wavelength, indicating that the optical rotatory angle of the solution perfectly compensates for the initial offset between the crossed polarizers. High-frequency oscillations observed at the ultraviolet and infrared boundaries are attributed to the amplification of thermal noise where the source intensity approaches zero.}
		\label{fig:spectra_conc}
	\end{figure*}
	
	\begin{figure*}[htbp]
		\centering
		\begin{subfigure}[b]{0.32\linewidth}
			\centering
			\includegraphics[width=\linewidth, height=4.5cm]{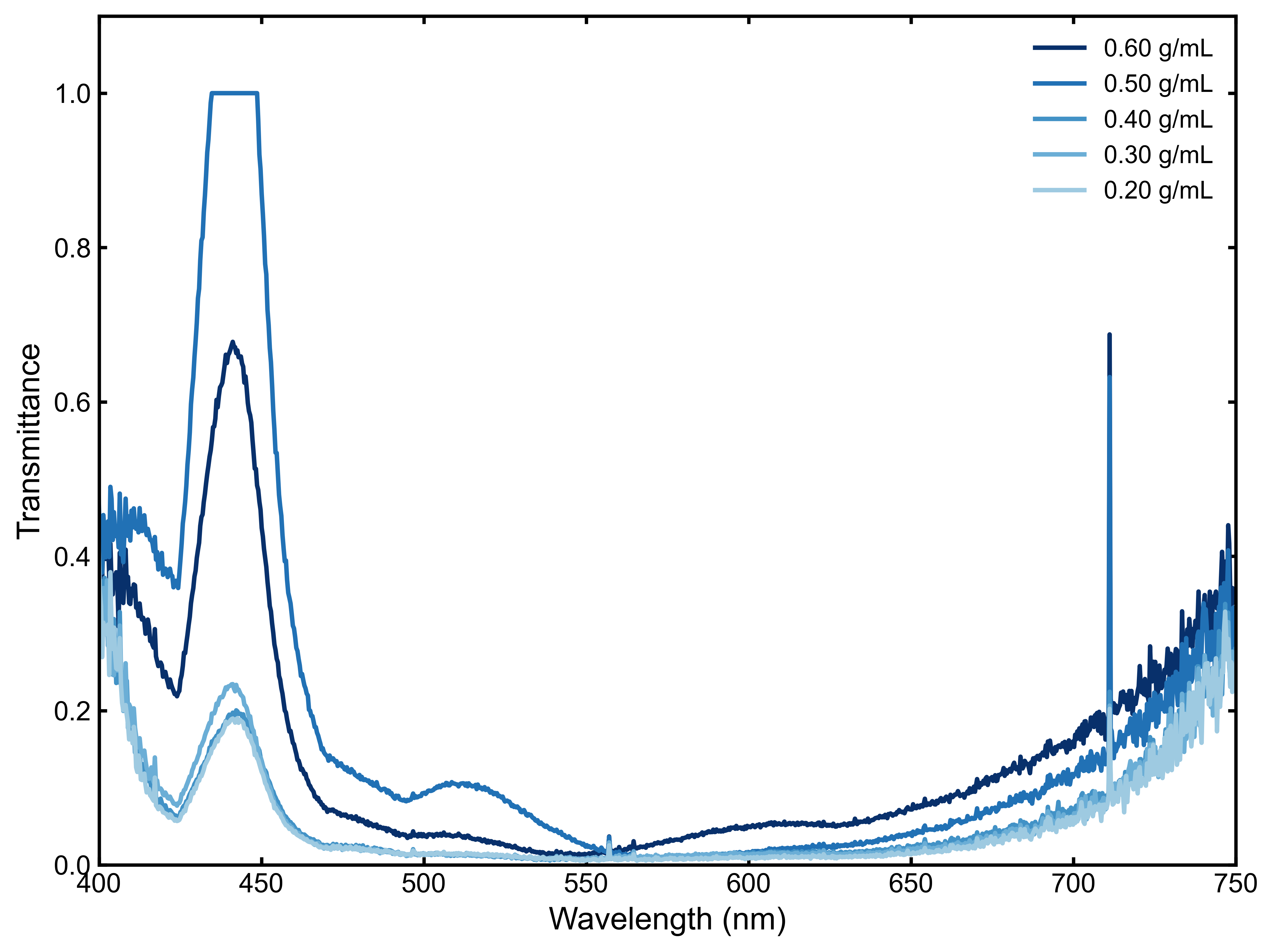}
			\caption{Concentration dependence}
			\label{fig:conc}
		\end{subfigure}\hfill
		\begin{subfigure}[b]{0.32\linewidth}
			\centering
			\includegraphics[width=\linewidth, height=4.5cm]{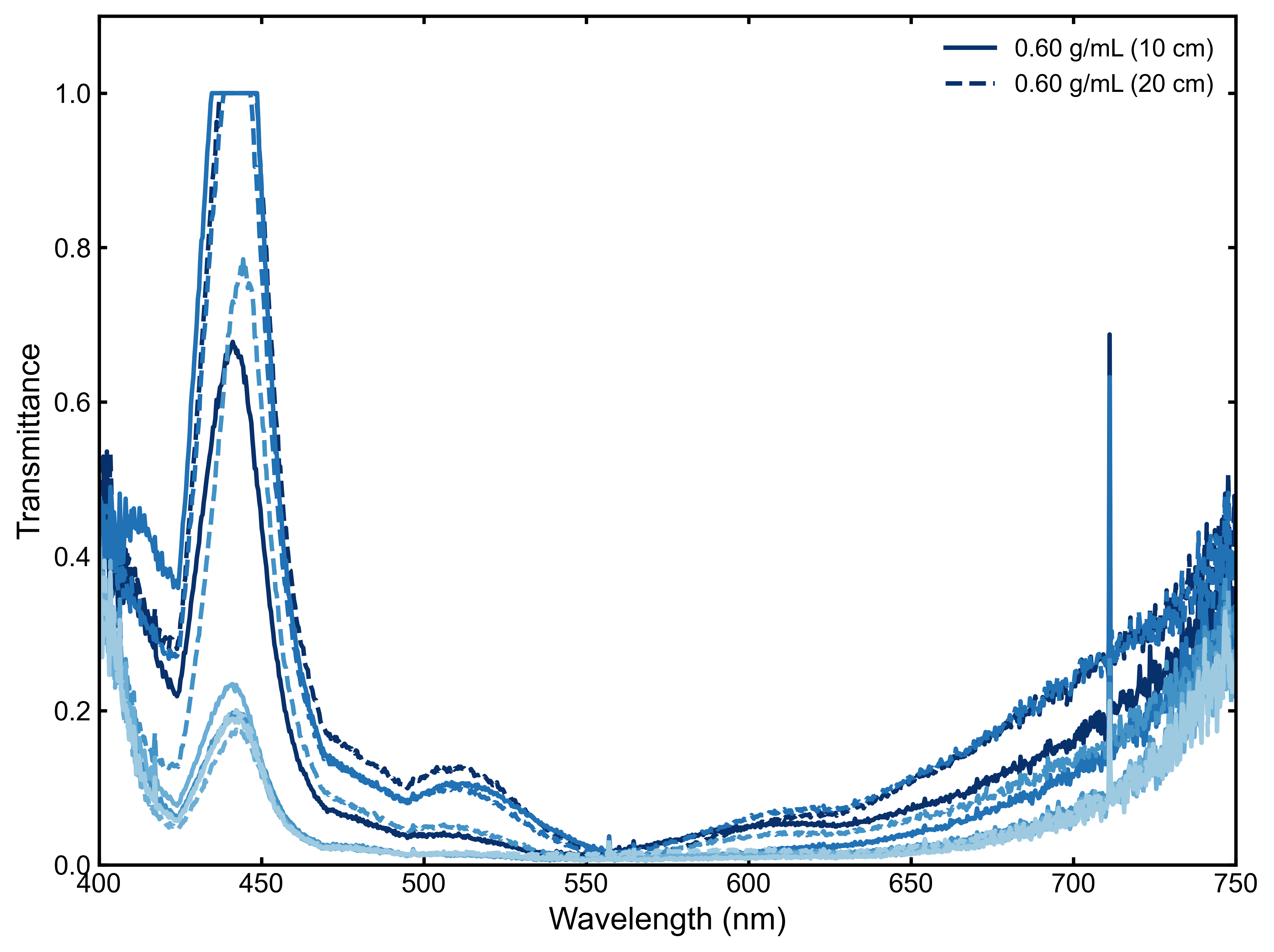}
			\caption{Path length amplification}
			\label{fig:path}
		\end{subfigure}\hfill
		\begin{subfigure}[b]{0.32\linewidth}
			\centering
			\includegraphics[width=\linewidth, height=4.5cm]{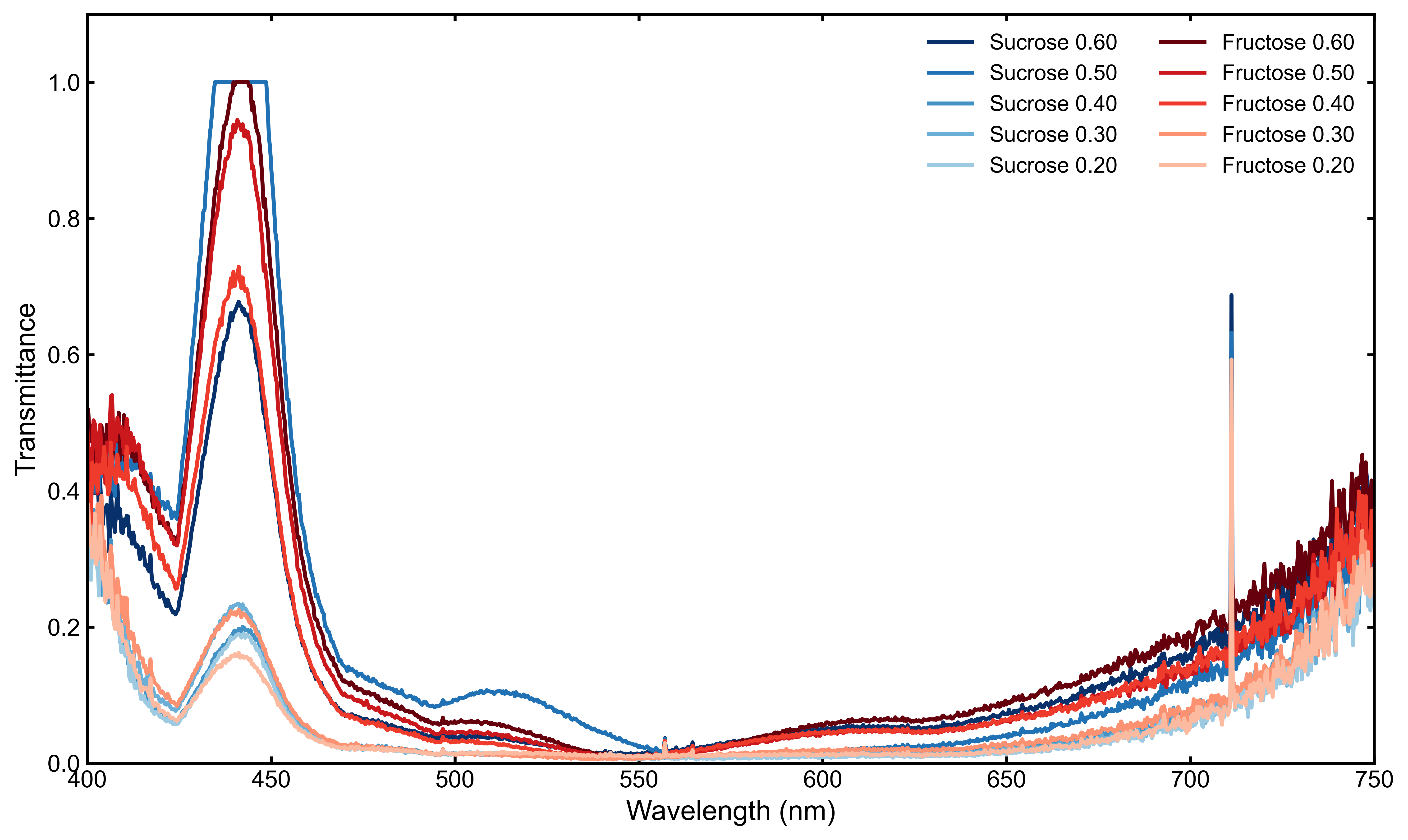}
			\caption{Chiral structural dependence}
			\label{fig:chiral}
		\end{subfigure}
		\caption{Multi-dimensional modulation of the normalized transmittance spectra. (a) Concentration dependence: The un-filtered spectra of sucrose at a fixed path length illustrate the continuous tuning of the transmission envelope via concentration gradients. A physical transmission limit is reached at $0.50$\,g/mL, where the specific rotatory angle perfectly compensates for the initial crossed-polarizer configuration. (b) Path length amplification: Utilizing smoothed data to eliminate high-frequency baseline noise, the cross-comparison between $10$\,cm (solid lines) and $20$\,cm (dashed lines) reveals that doubling the geometric propagation distance proportionally amplifies the phase retardation. This results in significant spectral shifts and modifications to the interference envelope's compression ratio. (c) Chiral structural dependence: Juxtaposing the spectra of dextrorotatory sucrose and levorotatory fructose under identical concentration protocols highlights the fundamental impact of the intrinsic specific rotation parameter ($K$). The distinct spectral positions and bandwidths between the two media rigorously confirm that chiral molecule selection serves as a macroscopic design parameter for defining the monochromator's active filtering band.}
		\label{fig:multidim}
	\end{figure*}
	
	To transcend qualitative visual observations, precise quantitative spectral characterization was performed to systematically validate the non-linear transmission mechanism governed by the ORD effect. The raw transmitted intensity profiles were continuously monitored across the predefined variable groups. These initial measurements, while clearly exhibiting distinct narrow-band transmission peaks corresponding to specific polarization matching conditions, were significantly convoluted by the non-flat emission baseline of the wideband halogen source and thermal noise. 
	By applying the established full-spectrum point-by-point normalization protocol, we successfully decoupled these instrumental artifacts, yielding the pure intrinsic transmittance spectra. This critical data transformation step isolated the periodic modulation induced exclusively by the chiral medium, enabling the accurate extraction of the full width at half maximum (FWHM) without background contamination. Subsequent analysis of these normalized interference envelopes under varying concentrations, path lengths, and chiral structures provides robust quantitative evidence for the system's spectral tuning capabilities.
	
	Within the full-spectrum normalized data, two non-physical anomalies necessitate error attribution. In the extreme short-wave ($<400$\,nm) and long-wave ($>750$\,nm) regions, the curves exhibit violent random oscillations, occasionally breaching the 1.0 theoretical limit. This numerical divergence occurs because the source radiation is negligible at these edges, driving the denominator $(I_{\text{blank}} - I_{\text{dark}})$ in the normalization equation toward zero and sharply amplifying high-frequency thermal noise. Additionally, sharp vertical spikes near 710\,nm and 780\,nm---invariant to solution concentration or type---were identified as localized hot pixels on the CCD array or non-uniform stray light leakage. To prevent these artifacts from corrupting automated bandwidth extraction, subsequent data processing was truncated to a valid confidence interval of 420--700\,nm and smoothed using a second-order Savitzky--Golay filter.
	
	Constrained by current laboratory conditions and non-ideal polarizers, the minimum output bandwidth achieved by this ORD monochromator remains on the order of 20\,nm. Although this metric---limited by component surface figures, leakage factor $d$, and stray light---diverges from the ideal sub-nanometer limit by three orders of magnitude, it precisely calibrates the true performance boundaries of such monochromators within a classical physical framework.
	
	\section{Error Analysis and Future Perspectives}
	\label{sec:error}
	
	The discrepancies between the macroscopically weighted theoretical ORD curves and experimental standards (\SI{5}{\percent}--\SI{18}{\percent} at \SI{589}{\nano\meter}) stem from intrinsic TD-DFT limitations, including systematic deviations in excited-state energies and implicit solvation model approximations. (Note that isolated conformers can show larger deviations, but these are largely averaged out in the Boltzmann-weighted ensemble.) Experimentally, the $\sim\SI{20}{\nano\meter}$ bandwidth limit arises from compounding macroscopic constraints: low photon flux of halogen lamps in the violet region, degraded extinction ratios of polymer polarizers at short wavelengths, residual impurities introducing non-rotatory absorption, and quantification uncertainty from camera-based pixel tracking.
	
	Future experimental architectures must upgrade the optical train: replacing broad-spectrum lamps with supercontinuum lasers or narrow-band LEDs, substituting polymer films with calcite Glan-Taylor prisms or high-extinction wire-grid polarizers, and utilizing decolorized ultra-pure samples with cooled CCD spectrometers for direct quantitative measurement of the transmission FWHM at the intensity minimum. Through these coordinated upgrades, the current gap between experiment and theoretical limit can be compressed to within a single order of magnitude.
	
	Beyond these classical improvements, this domain further encompasses advanced topics including single-photon detection, quantum-limited sensing, Variational Quantum Eigensolver (VQE) algorithms for quantum simulation, and quantum information processing. By housing the optical apparatus within a dilution refrigerator to maximally suppress thermal decoherence, the system can be effectively driven to the single-photon level. Under these extreme conditions, we project two primary operational scenarios: conventional classical light input resulting in a naturally filtered single-photon output, and a strict single-photon input to single-photon output continuous mapping. The rigorous theoretical derivations governing these non-classical dynamics, along with their specific application scenarios, constitute the core of our ongoing and future research endeavors.
	
	\section{Quantum Limit Regime}
	
	At ultralow temperatures ($\sim$10~mK), phonon-induced decoherence in solid-state chiral media or doped-ion systems is completely suppressed, allowing the homogeneous linewidth to reach the fundamental spontaneous emission limit. Under such extreme physical conditions, the frequency-resolving power of an optical rotatory dispersion filter is no longer bottlenecked by classical thermal broadening, but is fundamentally constrained by quantum photon shot noise and the finite heat capacity of the ultralow-temperature environment.
	
	\textbf{Macroscopic phenomenological dispersion model.---}Starting from a macroscopic single-absorption-band dispersion model, chiral molecules are treated as being frozen within a transparent solid matrix. When the system is biased at the optimal measurement point for maximum sensitivity, the ultimate limit of frequency estimation uncertainty for a single measurement is strictly bounded by the quantum Cram\'er-Rao bound. The quantum Fisher information $F_Q$ determined by the coherent-state probe reveals the core constraint relating the average total number of input photons $N_{\mathrm{in}}$, the average optical depth $\mathrm{OD}$, and the rotatory angle $\theta$:
	\begin{equation}
		F_Q = N_{\mathrm{in}} \cdot 2e^{-\mathrm{OD}} \left(\frac{d\omega}{d\theta}\right)^2
	\end{equation}
	To ensure the physical observability of the filter, its intrinsic physical bandwidth must be resolvable against the quantum noise. However, at ultralow temperatures, every photon absorbed by the medium is dissipated as heat, threatening the stability of the dilution refrigerator. Consequently, within a fixed integration time $\tau$, the total number of input photons is subjected to a strict thermodynamic constraint governed by the refrigerator's maximum cooling power $Q_{\mathrm{cool}}$ at 10~mK:
	\begin{equation}
		N_{\mathrm{in}} \cdot \hbar\omega_0 \leq Q_{\mathrm{cool}}\tau
	\end{equation}
	Combining this energy constraint with the shot-noise limit, we derive the ultimate minimum bandwidth of the monochromator under the macroscopic dispersion model, parameterized by the radiatively limited linewidth $\Gamma$ and the chiral anisotropy factor $g$:
	\begin{equation}
		\Delta\omega_{\mathrm{min}} = \frac{\pi\Gamma}{g \ln\left(\frac{2Q_{\mathrm{cool}}\tau}{\pi^2 \hbar\omega_0}\right)}
	\end{equation}
	This macroscopic result qualitatively exposes the fundamental difficulty of dispersive filters: due to the exponential decay of transmission intensity, compressing the bandwidth requires an exponentially growing input optical power, causing the device performance to rapidly hit the thermal dissipation ceiling of the cryogenic environment.
	
	\textbf{Microscopic asymmetric Jaynes-Cummings model.---}Nevertheless, the aforementioned macroscopic phenomenological model exhibits physical inconsistencies when approaching extreme asymptotic limits, as it artificially splices the dispersion slope of an ideal cavity with finite losses, thereby neglecting the underlying microscopic mechanisms. To establish a rigorously self-consistent microscopic physical boundary, we introduce an asymmetric dual-mode cavity electrodynamics architecture, employing the microscopic Jaynes-Cummings (JC) model to formally capture the symbiotic relationship between dispersion and dissipation. In this scheme, the right-handed circularly polarized mode is strongly coupled to a solid-state two-level system (e.g., a superconducting qubit), forming a standard single-mode JC model, whereas the left-handed mode serves as an ideal reference channel directly transmitted with a unity transmission coefficient.
	
	Based on microscopic input-output theory and fully retaining the spontaneous emission rate $\gamma$ from the atomic excited state to non-cavity modes, the optimal operating point of the system driven by a weak coherent continuous wave naturally locks onto the spectral peak generated by the vacuum Rabi splitting in the strong-coupling regime. At this optimal bias point, the maximum sensitivity function $S_{\mathrm{max}}$, which characterizes the signal-to-noise ratio figure of merit per input photon, can be rigorously expressed analytically as a function of the cavity dissipation rate $\kappa$ and the atomic loss rate $\gamma$:
	\begin{equation}
		S_{\mathrm{max}} = \frac{4\kappa}{(\kappa+\gamma)^2} \cdot \frac{1}{\sqrt{1 + \left(\frac{\kappa}{\kappa+\gamma}\right)^2}}
	\end{equation}
	Simultaneously, at the center of the Rabi splitting peak, the actual absorption probability of a single incident photon is rigorously determined as $(3-T_R^{\mathrm{peak}})/4$. Constrained by the ultralow-temperature cooling power $Q_{\mathrm{cool}}$ at 10~mK, energy conservation further restricts the maximum number of input photons allowed within the measurement period $\tau$:
	\begin{equation}
		N_{\mathrm{in}}^{\mathrm{max}} = \frac{4Q_{\mathrm{cool}}\tau}{(3-T_R^{\mathrm{peak}})\hbar\omega_0}
	\end{equation}
	Synthesizing this maximum photon number with the sensitivity function eliminates all logical discontinuities, rigorously yielding the closed-form expression for the quantum-limited minimum bandwidth of the monochromator under the asymmetric dual-mode JC model:
	\begin{equation}
		\Delta\omega_{\mathrm{min}} = \frac{(\kappa+\gamma)^2}{4\kappa} \sqrt{ \frac{(1+T_R^{\mathrm{peak}})(3-T_R^{\mathrm{peak}})}{T_R^{\mathrm{peak}}} } \cdot \frac{1}{\sqrt{N_{\mathrm{in}}^{\mathrm{max}}}}
	\end{equation}
	Substituting typical superconducting transmon qubit parameters (e.g., transition frequency $\omega_0/2\pi \approx 5$~GHz, spontaneous emission lifetime $T_1 \approx 50~\mu\mathrm{s}$, corresponding to $\gamma/2\pi \approx 3.2$~kHz) into this self-consistent model, with a cooling power of $1~\mu\mathrm{W}$ and a 1-second integration time, the theoretical limit of frequency resolution reaches an astonishing nanohertz level ($\Delta f_{\mathrm{min}} \approx 1.5 \times 10^{-6}$~Hz). This compellingly demonstrates that optical rotatory filtering mediated by a single-atom strongly coupled cavity can, in principle, completely surpass the resolution limits of traditional classical devices.
	
	\textbf{Applications and challenges in superconducting quantum circuits.---}Such a quantum-limited monochromator, possessing extreme frequency resolution on the nanohertz scale, exhibits tremendous application potential in scalable superconducting quantum computing. Particularly in the execution of syndrome measurement circuits for complex topological quantum error correction codes (such as the XZZX-type surface code or the generalized surface code), the hardware architecture typically involves densely packed neighboring transmon couplers and readout cavities. When performing high-density frequency-division multiplexed (FDM) simultaneous readouts on multiple qubits, out-of-band microwave pump crosstalk and the Purcell effect induced by residual intra-cavity photons can easily cause unintended decoherence in non-target qubits. If this asymmetric dual-mode JC filter is integrated into the control layer or long-range coupling layer of the chip, its ultra-narrow physical bandwidth and exceptionally steep dispersion slope can precisely block out-of-band microwave crosstalk while maintaining high-fidelity signal extraction under ultralow photon-number distributions, thereby providing near-perfect isolation and protection for fragile error-correction quantum states during syndrome measurements.
	
	Although this filter establishes an extremely high theoretical resolution boundary, its practical implementation in superconducting quantum chips must overcome numerous technical challenges. In an ultralow-temperature solid matrix, $1/f$ fluctuations induced by charge noise, minute substrate temperature instabilities, and external ambient flux noise all induce microscopic drifts in the artificial atom's resonance frequency $\omega_0$. These technical drifts can easily mask the fine spectral features at the nanohertz scale. To address this practical dilemma, future mitigation strategies must be multipronged: First, advanced dynamical decoupling pulse techniques should be introduced into the control sequences preceding the filter, resetting the system phase via fast spin-echo sequences to suppress the modulation of the pole frequency by low-frequency environmental noise. Second, the left-handed ideal reference channel, which is immune to absorption, should be fully utilized to extract real-time phase-drift error signals. This enables the construction of a sub-microsecond closed-loop negative feedback stabilization system via high-performance field-programmable gate arrays (FPGAs) to firmly lock the resonance frequencies of the cavity and the qubit. Third, from a hardware material and structural perspective, employing three-dimensional superconducting cavity architectures or geometrically optimizing the transmon capacitor plates can drastically reduce surface dielectric losses, allowing the physical linewidth to approach the radiation limit assumed in our derivation to the greatest extent possible.
	
	\textbf{Summary and discussion.---}Analyzing and comparing the two aforementioned derivation schemes clearly reveals how they delineate the quantum physical boundaries of monochromators at ultralow temperatures from distinct theoretical dimensions. The macroscopic single-absorption-band model, predicated on the classical Cram\'er-Rao bound and the phenomenological Lorentzian dispersion relation, offers the advantage of an intuitive physical picture. It directly establishes a straightforward connection between macroscopic optical parameters (such as the optical depth $\mathrm{OD}$ and the anisotropy factor $g$) and quantum shot noise. However, because this scheme intrinsically decouples the dissipation process from the dispersion slope, it inevitably introduces logical inconsistencies when treating extreme asymptotic limits.
	
	In contrast, the microscopic JC model based on the asymmetric dual-mode architecture exhibits profound physical rigor and self-consistency. By bridging microscopic input-output theory with cavity quantum electrodynamics, it directly weaves the atomic spontaneous emission loss $\gamma$ into the Hamiltonian as an irreducible intrinsic variable. This rigorously proves that the optimal operating point is not arbitrarily chosen, but naturally determined by the equilibrium between the group delay and the transmittance at the Rabi splitting peak.
	
	At a deeper level, these two schemes exhibit a fundamental disparity in their physical scaling laws. In the macroscopic model, the dependence of the bandwidth on the available photon number follows a logarithmic relation ($\Delta\omega \sim 1/\ln N$), implying that compressing the bandwidth by a factor of two incurs a quadratic power cost. Conversely, in the microscopic JC model, the introduction of the coherent enhancement effect within the strongly coupled resonant cavity optimizes the relationship between the bandwidth and the maximum input photon number into an algebraic power law ($\Delta\omega \sim 1/\sqrt{N}$). This fundamental paradigm shift in the scaling law is the precise theoretical origin of why superconducting qubit systems can transcend the limitations of classical dispersive devices and achieve an ultimate nanohertz-level frequency resolution. Therefore, while the macroscopic scheme provides macroscopic intuition for the energy-precision tradeoff, the microscopic JC scheme offers a microscopic blueprint---combining self-consistency and operational feasibility---for realizing ultimate metrological devices in the microwave regime. The synthesis and contrast of these two frameworks not only complete the theoretical chain of ultralow-temperature optical rotatory filters but also illuminate a clear trajectory for hardware optimization in future quantum metrology and precision detection technologies.
	
	\section{Conclusion}
	\label{sec:conclusion}
	
	In summary, we have established a comprehensive framework bridging classical optical filtering with the quantum limits of chiral light-matter interaction. Our TD-DFT calculations, validated by macroscopic experiments, reveal that the narrowest achievable bandwidth is fundamentally determined by the Cotton effect near the molecular absorption band. While current classical experiments yield a bandwidth of $\sim\SI{20}{\nano\meter}$, our analysis predicts that visible-absorbing chiral molecules can achieve sub-nanometer bandwidths. Extending this framework to the single-photon quantum limit at ultra-low temperatures establishes the concept of the ``quantum monochromator,'' where spectral purity is constrained solely by the Heisenberg uncertainty principle. This work provides first-principles criteria for the ultimate performance of chiral optical filters and opens new avenues for quantum chiral spectroscopy and information processing.
	
	\begin{acknowledgments}
		We gratefully acknowledge the School of Physics at Xidian University for providing the experimental apparatus and valuable guidance for the classical monochromator measurements. Computational resources were supported by the Jinghang Ruichuang Supercomputing Platform of Xidian University. We extend our appreciation to Zichen Guo for providing the data processing templates. Furthermore, M.X. acknowledges the data validation assistance from colleagues at Peking University and Southern University of Science and Technology.
	\end{acknowledgments}
	
	\appendix
	
	\section{Computational Methods for Optical Rotatory Dispersion}
	\label{app:methods}
	
	All DFT and TD-DFT calculations employed a hierarchical strategy to balance thermodynamic accuracy with excited-state precision. The SMD implicit solvation model (water) was consistently applied across all computational steps.
	
	\textbf{Geometry Optimization and Frequency Calculations.} Initial geometry optimizations and harmonic vibrational frequency calculations were conducted at the B3LYP/6-31+G(d) level. This combination was chosen because B3LYP is the industry standard for organic molecular geometries, while the diffuse functions in 6-31+G(d) are essential for accurately describing the lone-pair electron density on hydroxyl and ether oxygens, preventing the overestimation of intramolecular hydrogen bond strengths. Frequency calculations at the same level verified that all optimized structures are true minima on the potential energy surface and provided thermodynamic corrections (zero-point energy, thermal enthalpy, and entropy contributions).
	
	\textbf{Single-Point Energy Corrections.} On the B3LYP/6-31+G(d) optimized geometries, single-point electronic energies were evaluated using the long-range corrected, dispersion-corrected functional wB97XD in conjunction with the all-electron Karlsruhe basis set def2TZVP. The choice of wB97XD was motivated by two critical factors: (i) its asymptotic recovery of 100\% Hartree-Fock exchange correctly describes the long-range electrostatic interactions essential for hydrogen bonding; (ii) the Grimme D2 empirical dispersion correction explicitly accounts for the weak intramolecular van der Waals interactions that govern sugar conformational folding. The def2TZVP basis set provides balanced triple-zeta quality for valence electrons, offering sufficient flexibility to describe the polarization effects of oxygen lone pairs without the prohibitive cost of quadruple-zeta bases.
	
	\textbf{Gibbs Free Energy Synthesis.} The final Gibbs free energies used for deriving room-temperature (\SI{293}{\kelvin}) Boltzmann population weights were synthesized by combining the wB97XD/def2TZVP electronic energies with the B3LYP/6-31+G(d) thermodynamic corrections. This composite scheme allocates the highest computational resource to the term most sensitive to conformational energy differences---the electronic energy---while using a validated moderate level for the less method-sensitive thermal corrections.
	
	\textbf{High-Accuracy ORD Sweeps.} Frequency-dependent specific rotations were calculated using the long-range corrected functional cam-B3LYP combined with Dunning's correlation-consistent basis set aug-cc-pVTZ. The cam-B3LYP functional, which partitions the Coulomb operator into short- and long-range components, asymptotically recovers 100\% Hartree-Fock exchange. This is physically imperative for eliminating the systematic overestimation of charge-transfer and Rydberg-type excitation energies inherent in standard functionals like B3LYP. The aug-cc-pVTZ basis set provides diffuse functions on both heavy atoms and hydrogen, offering indispensable flexibility for describing the diffuse electron density of high-lying excited states that govern the far-UV virtual transitions critical to ORD frequency dependence. Sweep calculations were executed at five characteristic wavelengths (356, 436, 546, 589, and \SI{656}{\nano\meter}), corresponding to common laboratory spectral lines, to construct the complete ORD curves.

	\section{Quantum Fisher Information and Self-Consistent Boundary of the Macroscopic Single-Absorption-Band Dispersion Model}
	\label{app:macroscopic_fisher}
	
	This section details the mathematical derivation of the fundamental limit imposed by shot noise on the frequency resolution under the macroscopic single-absorption-band model.
	
	Based on the classical Lorentz model, near the resonance, the complex refractive index difference at frequency $\omega$ can be expressed as:
	\begin{equation}
		\Delta n(\omega) = \frac{S}{2(\omega_0 - \omega - i\Gamma/2)},
	\end{equation}
	where $S$ is the oscillator strength parameter, and $\Gamma$ denotes the full width at half maximum (linewidth) at the spontaneous emission limit. Expanding this expression, its real part (representing optical rotatory dispersion) and imaginary part (representing circular dichroism) are respectively:
	\begin{equation}
		\Delta n'(\omega) = \frac{S}{2} \frac{\omega_0 - \omega}{(\omega_0 - \omega)^2 + (\Gamma/2)^2},
	\end{equation}
	\begin{equation}
		\Delta \kappa(\omega) = \frac{S}{2} \frac{\Gamma/2}{(\omega_0 - \omega)^2 + (\Gamma/2)^2}.
	\end{equation}
	
	The circular dichroism absorption coefficient difference is defined as $\Delta \alpha(\omega) = (2\omega/c) \Delta \kappa(\omega)$. At the resonance center $\omega = \omega_0$, this value reaches its maximum $\Delta \alpha_{\mathrm{max}}$, from which the oscillator strength parameter can be inversely derived:
	\begin{equation}
		S = \frac{c\Gamma\Delta\alpha_{\mathrm{max}}}{\omega_0}.
	\end{equation}
	The macroscopic rotatory angle of the medium is defined as $\theta(\omega) = (\omega L / 2c) \Delta n'(\omega)$. Taking the derivative with respect to frequency $\omega$, the dispersion slope reaches its maximum at the resonance center $\omega_0$, since the real part of the complex refractive index difference crosses zero:
	\begin{equation}
		\left.\frac{d\theta}{d\omega}\right|_{\omega_0} = \frac{L \Delta \alpha_{\mathrm{max}}}{2\Gamma}.
	\end{equation}
	
	When the filter is utilized as a spectral analyzer, the analyzer is biased at the point of maximum sensitivity (i.e., $\theta - \phi = \pi/4$). For a small frequency detuning, its intensity transmittance can be approximately expanded as:
	\begin{equation}
		T(\omega) \approx \frac{1}{2} e^{-\mathrm{OD}} \left[ 1 - \sin\left( 2 \frac{d\theta}{d\omega} (\omega - \omega_0) \right) \right].
	\end{equation}
	In this pure-loss channel, the quantum Fisher information determined by a coherent-state probe is equivalent to the classical Fisher information, derived directly from the transmission function:
	\begin{equation}
		\mathcal{F}_Q = \frac{N_{\mathrm{in}}}{T(\omega)} \left( \frac{\partial T}{\partial \omega} \right)^2.
	\end{equation}
	
	Substituting the bias point parameters $T = \frac{1}{2}e^{-\mathrm{OD}}$ and $\partial T / \partial \omega = e^{-\mathrm{OD}} |d\theta/d\omega|$, we obtain:
	\begin{equation}
		\mathcal{F}_Q = N_{\mathrm{in}} \cdot 2e^{-\mathrm{OD}} \left( \frac{d\theta}{d\omega} \right)^2.
	\end{equation}
	
	Utilizing the quantum Cram\'er-Rao bound, the ultimate limit of frequency estimation uncertainty for a single measurement is $\delta\omega_{\mathrm{min}} = 1/\sqrt{\mathcal{F}_Q}$. To ensure the physical observability of the filter, the monochromator's physical bandwidth itself must be resolvable against the quantum noise, necessitating the asymptotic self-consistent condition $\Delta\omega_{\mathrm{FWHM}} = \delta\omega_{\mathrm{min}}$. Combining this with the classical filter bandwidth definition $\Delta\omega_{\mathrm{FWHM}} = \pi \Gamma / (g \cdot \mathrm{OD})$ (where the anisotropy factor $g = \Delta\alpha_{\mathrm{max}} / \alpha_{\mathrm{avg}}$), we obtain the self-consistent equation:
	\begin{equation}
		\frac{\pi \Gamma}{g \cdot \mathrm{OD}} = \frac{2 \Gamma}{g \cdot \mathrm{OD} \cdot N_{\mathrm{in}} e^{-\mathrm{OD}}}.
	\end{equation}
	Simplifying this relation yields the fundamental constraint on the photon flux:
	\begin{equation}
		N_{\mathrm{in}} e^{-\mathrm{OD}} = \frac{2}{\pi^2} \approx 0.2026.
	\end{equation}

	\section{Microscopic Self-Consistent Derivation of the Asymmetric Dual-Mode Jaynes-Cummings Model}
	\label{app:microscopic_jc}
	
	This section aims to eliminate the logical discontinuity inherent in the ideal cavity assumption from a microscopic perspective, rigorously deriving the optimal bias point and dispersion slope by formally incorporating the atomic spontaneous emission rate $\gamma$.
	
	Considering an asymmetric dual-mode architecture, the right-handed (R) circularly polarized mode is strongly coupled to a two-level atom, whereas the left-handed (L) mode serves as an ideal reference channel with a unity transmission coefficient. In the rotating frame, the Hamiltonian of the right-handed mode is described by:
	\begin{equation}
		\hat{H} = \omega_c \hat{a}_R^\dagger \hat{a}_R + \frac{\omega_a}{2}\hat{\sigma}_z + g_0(\hat{a}_R \hat{\sigma}_+ + \hat{a}_R^\dagger \hat{\sigma}_-).
	\end{equation}
	Under a weak coherent continuous-wave drive, applying microscopic input-output theory yields the linear steady-state transmission coefficient for the right-handed mode:
	\begin{equation}
		t_R(\omega) = \frac{\kappa/2}{i(\omega_c - \omega) + \kappa/2 + \chi(\omega)},
	\end{equation}
	where the self-energy term $\chi(\omega)$ induced by the two-level atom is:
	\begin{equation}
		\chi(\omega) = \frac{g_0^2}{i(\omega_a - \omega) + \gamma/2}.
	\end{equation}
	
	In the strong-coupling regime ($g_0 \gg \kappa, \gamma$), the transmission spectrum exhibits vacuum Rabi splitting. The splitting peak centers are approximately located at:
	\begin{equation}
		\omega_{\pm} \approx \omega_0 \pm \sqrt{g_0^2 - \frac{\gamma^2}{4}}.
	\end{equation}
	At these resonance peaks $\omega_{\pm}$, the imaginary part of the self-energy vanishes. The pole condition simplifies the transmission coefficient to $t_R(\omega_{\pm}) \approx \kappa / (\kappa + \gamma)$. Consequently, the intensity transmittance at the peak is rigorously determined as:
	\begin{equation}
		T_R^{\mathrm{peak}} = \left(\frac{\kappa}{\kappa + \gamma}\right)^2.
	\end{equation}
	
	Taking the derivative of the phase $\phi_R(\omega) = \arg t_R(\omega)$ yields the dispersion slope. Near the pole of the Lorentzian splitting peak, its half-width $\Gamma_{\mathrm{mode}} \approx (\kappa + \gamma)/2$ dictates that the maximum phase slope is:
	\begin{equation}
		\left| \frac{d\phi_R}{d\omega} \right|_{\mathrm{max}} \approx \frac{4}{\kappa + \gamma}.
	\end{equation}
	Since the specific rotatory angle is defined as $\theta(\omega) = -\frac{1}{2}\phi_R(\omega)$, the maximum dispersion slope of the rotation angle becomes:
	\begin{equation}
		\left| \frac{d\theta}{d\omega} \right|_{\mathrm{max}} \approx \frac{2}{\kappa + \gamma}.
	\end{equation}
	
	Substituting the derived transmittance and dispersion slope into the universal sensitivity function $S(\omega) = \frac{2\sqrt{T_R}}{\sqrt{1+T_R}}\left|\frac{d\theta}{d\omega}\right|$, we obtain the optimal sensitivity at the center of the Rabi splitting peak:
	\begin{equation}
		S_{\mathrm{max}} = \frac{2\left(\frac{\kappa}{\kappa+\gamma}\right)}{\sqrt{1+\left(\frac{\kappa}{\kappa+\gamma}\right)^2}} \cdot \frac{2}{\kappa + \gamma}.
	\end{equation}
	Simplifying this expression rigidly yields the self-consistent unified limit formula:
	\begin{equation}
		S_{\mathrm{max}} = \frac{4\kappa}{(\kappa+\gamma)^2} \cdot \frac{1}{\sqrt{1 + \left(\frac{\kappa}{\kappa+\gamma}\right)^2}}.
	\end{equation}
	
\section{Thermodynamic Data, Specific Rotations, and Coordinates}
\label{app:data}

Table~\ref{tab:specific_rotation} compares the calculated and experimental specific rotations at 589 nm for various saccharide conformations, validating our computational methodology. Furthermore, Table~\ref{tab:gibbs} summarizes the absolute electronic energies ($E_{\text{wB97XD}}$), thermodynamic corrections ($G_{\text{corr}}$), final Gibbs free energies ($G$), and relative free energies ($\Delta G$) for all investigated saccharide conformations. Table~\ref{tab:boltzmann} details the derived Boltzmann population fractions at \SI{293}{\kelvin}. The Cartesian coordinates (in XYZ format) for all optimized geometries are provided as separate data files attached to this supplementary material.



\newpage

\begin{table}[htbp]
	\centering
	\caption{Comparison of Calculated and Experimental Specific Rotations at 589 nm for Saccharide Conformations}
	\label{tab:specific_rotation}
	\begin{tabular}{lcccc}
		\toprule
		\textbf{Conformation} & \textbf{Calculated ($^\circ$)} & \textbf{Experimental ($^\circ$)} & \textbf{Difference ($^\circ$)} & \textbf{Remarks} \\
		\midrule
		$\beta$-D-Fructopyranose  & -153.59 & $\sim$ -132.0 & -21.59 & Initial crystal \\
		$\alpha$-D-Glucopyranose  & +101.20 & $\sim$ +112.0 & -10.80 & Initial crystal \\
		$\beta$-D-Fructofuranose  & -87.90  & $\sim$ -78.0  & -9.90  & - \\
		$\beta$-D-Glucopyranose   & +66.81  & $\sim$ +52.5  & +14.31 & Aqueous equilibrium \\
		$\alpha$-D-Fructopyranose & -66.63  & $\sim$ -63.6  & -3.03  & - \\
		$\alpha$-D-Fructofuranose & -33.71  & $\sim$ -21.0  & -12.71 & Literature value disputed \\
		\midrule
		\textbf{Sucrose}          & +54.40  & $\sim$ +66.5  & -12.10 & Aqueous solution \\
		\bottomrule
	\end{tabular}
	\vspace{1ex}
	\raggedright
	\small{\textit{Note:} Calculations were performed at the CAM-B3LYP/aug-cc-pVTZ level in conjunction with the SMD (water) implicit solvation model. The difference is defined as $\Delta[\alpha] = [\alpha]_{\text{calc}} - [\alpha]_{\text{exp}}$.}
\end{table}

\begin{table}[htbp]
	\centering
	\caption{Gibbs free energies and relative energies of saccharide conformations calculated at the wB97XD/def2TZVP//B3LYP/6-31+G(d) level with SMD solvent model (water).}
	\label{tab:gibbs}
	\begin{tabular}{lcc}
		\toprule
		\textbf{Conformation} & \textbf{\(G\) (Hartree)} & \textbf{\(\Delta G\) (kcal/mol)} \\
		\midrule
		$\beta$-D-fructopyranose    & -687.072504 & 0.000 \\
		$\beta$-D-fructofuranose    & -687.071925 & 0.363 \\
		$\alpha$-D-fructofuranose   & -687.069425 & 1.932 \\
		$\alpha$-D-fructopyranose   & -687.068089 & 2.770 \\
		$\beta$-D-glucopyranose     & -687.071187 & 0.000 \\
		$\alpha$-D-glucopyranose    & -687.071085 & 0.064 \\
		\bottomrule
	\end{tabular}
	\vspace{1ex}
	\raggedright
	\small{\textit{Note:} The relative free energy for glucose is defined with respect to $\beta$-D-glucopyranose as the reference.}
\end{table}

\begin{table}[htbp]
	\centering
	\caption{Boltzmann population fractions of saccharide conformations at \SI{293}{\kelvin}.}
	\label{tab:boltzmann}
	\begin{tabular}{lcc}
		\toprule
		\textbf{Sugar} & \textbf{Conformation} & \textbf{Population (\%)} \\
		\midrule
		Fructose & $\beta$-D-fructopyranose  & 62.92 \\
		Fructose & $\beta$-D-fructofuranose  & 34.08 \\
		Fructose & $\alpha$-D-fructofuranose & 2.41 \\
		Fructose & $\alpha$-D-fructopyranose & 0.59 \\
		Glucose  & $\beta$-D-glucopyranose   & 52.7 \\
		Glucose  & $\alpha$-D-glucopyranose  & 47.3 \\
		\bottomrule
	\end{tabular}
\end{table}

\end{document}